\documentclass[twocolumn,10pt]{asme2ej}

\usepackage{epsfig} 

\title{Backlash Identification in Two-Mass Systems by Delayed Relay Feedback}

\author{Michael Ruderman \thanks{All correspondence should be addressed to this author.}
    \affiliation{
    University of Agder\\
    4879 Grimstad, Norway\\
    Email: michael.ruderman@uia.no
    }
}

\author{Shota Yamada, Hiroshi Fujimoto
    \affiliation{ The University of Tokyo\\
    277-8561 Chiba, Japan\\
    Email: yamada.shota13@ae.k.u-tokyo.ac.jp, fujimoto@k.u-tokyo.ac.jp
    }
}

\begin{document}

\maketitle

\begin{abstract}
{\it Backlash, also known as mechanical play, is a piecewise
differentiable nonlinearity which exists in several actuated
systems, comprising, e.g., rack-and-pinion drives, shaft
couplings, toothed gears, and other machine elements. Generally,
the backlash is nested between the moving parts of a complex
dynamic system, which handicaps its proper detection and
identification. A classical example is the two-mass system which
can approximate numerous mechanisms connected by a shaft (or link)
with relatively high stiffness and backlash in series. Information
about the presence and extent of the backlash is seldom exactly
known and is rather conditional upon factors such as wear, fatigue
and incipient failures in the components. This paper proposes a
novel backlash identification method using one-side sensing of a
two-mass system. The method is based on the delayed relay operator
in feedback that allows stable and controllable limit cycles to be
induced and operated within the (unknown) backlash gap. The system
model, with structural transformations required for the one-side
backlash measurements, is given, along with the analysis of the
delayed relay in velocity feedback. Experimental evaluations are
shown for a two-inertia motor bench that has coupling with
backlash gap of about one degree. }
\end{abstract}

\section{Introduction}
\label{sec:1}

Mechanical backlash, i.e., the phenomenon of play between adjacent
movable parts, is well known and causes rather disturbing
side-effects such as lost motion, undesired limit cycles in a
closed control loop, reduction of the apparent natural frequencies
and other. For a review of this phenomenon we refer to
\cite{nordin2002}. It is the progressive wear and fatigue-related
cracks in mechanical structures that can develop over the
operation time of an actuated system with backlash; this is in
addition to appearance of a parasitic noise \cite{dion2009}.
Moreover, increasing backlash leads to more strongly pronounced
chaotic behavior \cite{tjahjowidodo2007} that, in general,
mitigates against accurate motion control of the system.

\begin{figure}[!h]
\centering
\includegraphics[width=0.75\columnwidth]{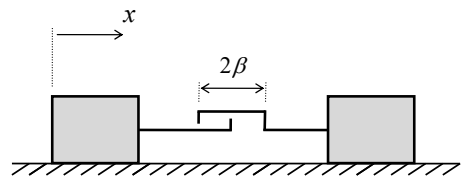}
\caption{Two-mass system with backlash.} \label{fig:1}
\end{figure}

Two-mass systems with backlash, as schematically shown in Fig.
\ref{fig:1}, constitute a rather large class of the motion
systems. Here, a driving member, which is a motor or generally
actuator, is coupled to the driven member (load) by various
construction elements, i.e., gears, couplings, or kinematic pairs.
The connecting elements contain a finite gap, here denoted by
$2\beta$, usually of a small size comparing to the rated relative
displacement of the motion system. Within this gap, both moving
masses become decoupled from each other. It is important to note
that this structure has a backlash in feedback which differs
significantly from feedthrough systems where a static backlash
element appears either in the input or output channel. Those types
of systems with backlash have also attracted considerable
attention in systems analysis and control, see, e.g.,
\cite{tao1995}, \cite{hagglund2007}. However, they should not be
confused with the types of backlash feedback systems as depicted
in Fig. \ref{fig:1} and addressed in the following. Some previous
studies which consider analysis and control of backlash feedback
systems can be found in, e.g.,
\cite{brandenburg1986,hori1994,gerdes1995impact}.

Despite the fact that the fundamental understanding of backlash
mechanisms appears to be something of a solved problem,
case-specific modeling, and above all, reliable detection and
identification still remain relevant and open-ended problems in
systems and control engineering. Applications can be found in
industrial robotic manipulators, e.g., \cite{ruderman2009},
flexible medical robots, e.g., \cite{morimoto2017}, servo drive
systems, e.g., \cite{gebler1998,villwock2009}, automotive power
trains, e.g., \cite{lagerberg2007} and other areas. Information
about the presence and extent of backlash, which is a rather
undesirable structural element, is seldom exactly known, and
usually requires identification under regular operational
conditions, i.e. during exploitation. When both the motor-side and
load-side of a two-mass (two-inertia) system are equipped with
high-resolution position sensors (encoders), see for instance
\cite{Yamada16}, the identification of backlash becomes a trivial
task which can be accomplished under various quasi-static,
low-excitation conditions. A frequently encountered case, however,
is when load-side sensing is not available, or is associated with
special auxiliary measures which can markedly differ from the
normal operational conditions. A typical situation occurs when
only the driving actuator is equipped with an encoder or other
position-giving sensor, while the total drive train, presumably
with backlash, has no access to additional measurements.

In the last two decades, various strategies for estimation, and
hence identification, of the backlash between two moving parts,
have been developed. A very early study and analysis of the
backlash phenomenon in geared mechanisms was made by
\cite{Tustin47}. A specially designed impulse excitation and motor
current filtering method was proposed in \cite{gebler1998} for
high-precision servo systems. While straightforward in
realization, the method relies on the assumption that the torque
impulses are sufficient to move the motor and, at the same time,
to keep the load immobile. For systems with high and uncertain
damping and additional elasticity modes, such excitation can be
challenging with regard to realization and manifestation of the
backlash. The nonlinear observation of backlash states was
proposed and evaluated in \cite{merzouki2007}, though both
motor-side and load-side measurements were required. An
identification approach based on the ridges and skeletons of
wavelet transforms was demonstrated by \cite{tjahjowidodo2007b}.
The method requires a relatively broadband frequency excitation
and a combined time-frequency analysis that can be challenging in
a practical application. One of the most established strategies of
backlash identification, using motor-side sensing only, was
reported in \cite{villwock2009}. It should be noted, however, that
the underlying ideas were previously provided in \cite{Tustin47}.
This approach will also be taken as a reference method for
experimental evaluation in the present work. Another, more
frequency-domain-related approach for analyzing backlash by
approximating it with a describing function, can be found in
\cite{lichtsinder2016}. This relies on the so-called "exact
backlash mode" introduced in \cite{nordin1997new}. Describing
function analysis of systems with impact and backlash can be found
in \cite{barbosa2002}. An experimental comparison of several
backlash identification methods, mainly based on the previous
works of \cite{gebler1998,lagerberg2007,villwock2009}, can be
found in \cite{yang2012study}. There, the authors concluded that
the method of integration of the motor speed \cite{villwock2009}
was the most accurate from those under evaluation.

The objective of this paper is to introduce a new strategy for
identifying the backlash in two-mass systems when only a
motor-side sensing is available. In addition, this method does not
require a large excitation of the overall motion dynamics, in
contrast to \cite{villwock2009}. This can be advantageous for
various machines and mechanisms where the load cannot be driven at
higher velocities and accelerations. The proposed method relies on
the appearance of stable and controllable limit cycles, while
using a delayed relay in the velocity feedback loop. We recall
that relay feedback systems have been intensively studied since
the earlier pioneering works \cite{andronov1949,mullin1959} and
successfully applied in, e.g., auto-tuning of controllers
\cite{hang2002} and other purposes such as mass identification
\cite{mizuno2008}. In the present study, the uncertainties and
parasitic effects of the system dynamics are not explicitly taken
into account, but the relay in feedback provides the necessary
robustification, which is also confirmed by experiments. Further
we note that attempts to use a relay feedback for identifying
backlash in two-mass systems were also made in
\cite{han2016backlash}. However, the relay feedback was solely
used for inducing a large-amplitude periodic motion, while the
underlying identification strategy relied on the motor speed
pattern analysis, similar as in \cite{villwock2009}. Furthermore,
an exact detection of zero-crossing and extremum instants is
required, along with the ratio of the masses (correspondingly
inertias).

The rest of the paper is organized as follows. In section II we
describe the underlying modeling approach for two-mass systems
with backlash, and the required structural transformations which
allow consideration of the two-mass system as a plant with the
motor-side sensing only. Section III contains a detailed analysis
of the feedback relay system and induced limit cycles, including
controllable drifting, which enables operating the motor within
and beyond the backlash gap. An experimental case study with the
minimal necessary identification of system parameters and
relay-based backlash identification is provided in section IV,
together with comparison to the reference method reported in
\cite{villwock2009}. The paper is summarized and the conclusions
are drawn in section V.

\section{Two-Mass System with Backlash}
\label{sec:2}

\begin{figure}[!h]
\centering
\includegraphics[width=0.85\columnwidth]{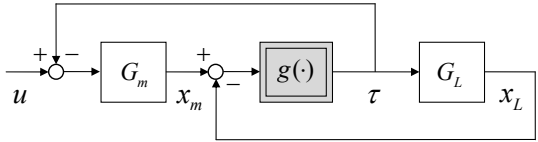}
\caption{Block diagram of two-mass system with backlash in the
link.} \label{fig:2}
\end{figure}

The block diagram of a generalized two-mass system with backlash
is shown in Fig. \ref{fig:2}. Note that this complies with the
principal structure introduced in section \ref{sec:1}, cf. with
Fig. \ref{fig:1}. The motor-side and load-side dynamics, described
in the relative coordinates $x_m$ and $x_L$ respectively, are
given by
\begin{eqnarray}
\label{eq:1}
  m \ddot{x}_m + d \dot{x}_m + f \mathrm{sign}(\dot{x}_m) = u- \tau, \\
  M \ddot{x}_L + D \dot{x}_L + F \mathrm{sign}(\dot{x}_L) = \tau.
\label{eq:2}
\end{eqnarray}
Both are coupled in the forward and feedback manner by the overall
transmitted force $\tau$, which is often referred to as a link (or
joint) force. Note that in the following we will use the single
terms \emph{force}, \emph{displacement} and \emph{velocity} while
keeping in mind the generalized forces, correspondingly
generalized motion variables. Thus, the rotary coordinates and
corresponding rotary degrees of freedom will be equally considered
without changing the notation of the introduced variables. For
example, in the case study of the two-inertia system presented in
section \ref{sec:4}. Clearly, the parameters $m$, $M$, $d$, $D$,
$f$ and $F$ are the motor and load masses, damping and Coulomb
friction coefficients respectively. We will also denote the known
invertible mapping $(u-\tau) \mapsto x_m$ by $G_m$ and $\tau
\mapsto x_L$ by $G_L$. Here, we explicitly avoid any notations of
the transfer function since the dynamics of (\ref{eq:1}) and
(\ref{eq:2}) include also the nonlinear terms of Coulomb friction.
Furthermore we note that despite nonlinear friction can impose
more complex by-effects, see e.g. \cite{ruderman2015,ruder2017}, a
constant Coulomb friction only is considered here. This is
justified by using the discontinuous relay operator in feedback
which allows overcoming stiction and other transient by-effects of
the nonlinear friction. Using the nonlinear function $g(\cdot)$,
where the parameters are unknown but a certain structure can be
assumed, we will elucidate the force transmission characteristics
of the link.

Usually the link is excited by the relative displacement, directly
resulting from
\begin{equation}
\delta = x_m - x_L, \label{eq:3}
\end{equation}
between the motor and load. We assume that the link transfer
characteristics offer a relatively high, but yet finite, stiffness
and incorporate a backlash connected in series. Under high
stiffness conditions we understand the first resonance peak which
can be detected and identified within the measurable frequency
range, however sufficiently beyond the input frequencies and
control bandwidth of the system.

We assume that the backlash itself is sufficiently damped so that
the relative displacement between the "impact pair" is not subject
to any long-term oscillations when the mechanics engage. Here, we
recall that the simplest modeling approach for the backlash in
two-mass systems, used for instance in
\cite{villwock2009,Yamada16}, applies the dead-zone operator (with
$\delta$ argument) only, while the dead-zone output is
subsequently gained by the stiffness of the link. This arrangement
results in a proper backlash behavior merely at lower frequencies,
but tends to produce spurious $\delta$-oscillations at higher
harmonics. More detailed consideration of the backlash in two-mass
systems incorporates a nonlinear damping at impact
\cite{gerdes1995impact}, for which advanced impact models can be
found, e.g., in \cite{hunt1975,lankarani1994}. Also note that in
\cite{nordin1997new}, an inelastic impact was considered when the
backlash gap was closed, and the switching distinction for gap and
contact modes was made, cf. further with eq. (\ref{eq:4}).

\begin{figure}[!h]
\centering
\includegraphics[width=0.8\columnwidth]{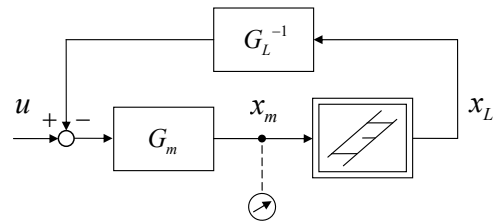}
\caption{Equivalent block diagram for modeling the two-mass system
with backlash described by the play-type hysteresis operator.}
\label{fig:3}
\end{figure}

The above assumption of sufficient backlash damping allows a
simple structural transformation of the block diagram from Fig.
\ref{fig:2} to be made. Assuming a rigid coupling at the backlash
contacts and loss-of-force transmission at the backlash gap, an
equivalent model can be obtained, as shown in Fig. \ref{fig:3}.
Obviously, the consideration of the backlash element in Fig.
\ref{fig:3} is purely kinematic, so that an ideal play-type
hysteresis \cite{visintin1994,Krejci96} is assumed between the
motor and load displacements. The kinematic backlash, which is the
Prandtl-Ishlinskii operator of the play type, can be written in
the differential form as
\begin{equation}\label{eq:4}
    \dot{x}_L = \left\{%
\begin{array}{ll}
    \dot{x}_m & \hbox{if } \; x_L = x_m - \beta, \; \dot{x}_m  > 0 \:,\\[0.1cm]
    \dot{x}_m & \hbox{if } \; x_L = x_m + \beta, \; \dot{x}_m  < 0 \:,\\[0.1cm]
    0 & \hbox{if } \; x_m - \beta < x_L < x_m + \beta \:,\\[0.1cm]
    0 & \hbox{otherwise} \:,
\end{array}%
\right.
\end{equation}
cf. with \cite{ruderman2009,ruderman2017}, where the total gap is
$2\beta$. Note that the backlash acts as a structure-switching
nonlinearity $x_L = \mathrm{P}[x_m]$, so that two operational
modes are distinguished. The first mode, within the backlash
\emph{gap} i.e. when $|\delta| < \beta$, implies zero feedback of
the $G_L^{-1}$ dynamics. The second mode, is during mechanics
\emph{engagement}, i.e., $|\delta| = \beta$ provides the feedback
coupling by $G_L^{-1}$, so that the system performs as a single
mass with the lumped parameters $(m+M)$ and $(d+D)$. During the
switching between both modes the play operator (\ref{eq:4})
becomes non-differentiable. That is, the first time-derivative of
the $\mathrm{P}$-output to an inherently $\mathcal{C}^2$-smooth
motor position input contains step discontinuities.
Correspondingly, the second time-derivative constitutes the
weighted delta impulses, in terms of the distribution theory.
These impulses can be interpreted as an instantaneous impact force
$\tau$ which excites the motor dynamics when the backlash switches
between gap and engagement, and vice versa. We should stress that
this switching mechanism does not explicitly account for the link
stiffness and damping at impact, as previously mentioned. However,
this is not critical at lower frequencies, since the link
stiffness is assumed to be sufficiently high, and the motor and
load dynamics are subject to the viscous and Coulomb friction
damping; cf. (\ref{eq:1}) and (\ref{eq:2}). Also we note that the
hybrid approach, with a switching structure for the gap and
engagement, was also used in \cite{rostalski2007} for modeling and
control of mechanical systems with backlash.

A significant feature of the equivalent model shown in Fig.
\ref{fig:3} is that it allows the two-mass system with backlash to
be considered as a single closed loop, while assuming that the
motor-side sensing only is available. This is a decisive property
we will make use of when identifying backlash by using one-side
sensing of the two-mass system. It is to recall that the loop is
closed during the \emph{engagement} mode and open during the
\emph{gap} mode of the backlash in dynamic system. This results in
an impulsive behavior during mode switching. Therefore, the motor
dynamics are disturbed periodically when a controllable stable
limit cycle occurs. We analyze this situation further in section
\ref{sec:3:sub:3}.

\section{Delayed Relay Feedback}
\label{sec:3}

\begin{figure}[!h]
\centering
\includegraphics[width=0.5\columnwidth]{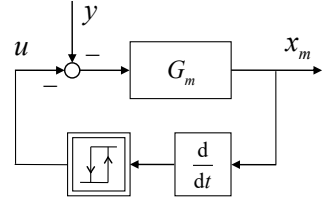}
\caption{Relay feedback system with motor dynamics in the loop.}
\label{fig:4}
\end{figure}

The proposed backlash identification is based on the delayed relay
in the feedback of the motor velocity. Consider the relay feedback
system as shown in Fig. \ref{fig:4}. Note that the structure
represents our system, as introduced in section \ref{sec:2}, when
operating in the backlash gap mode, i.e., $y=0$. From the point of
view of a control loop with relay, the switching feedback force
(see Fig. \ref{fig:3}) can be seen as an exogenous signal $y$. The
non-ideal relay, also known as hysteron \cite{visintin1994},
switches between two output values $\pm h$, while the switching
input is "delayed" by the predefined threshold values $\pm e$.
Therefore $2e$ is the relay width, corresponding to the hysteron's
size, and $h$ is the amplification gain of the relay output. The
hysteron with input $z$ is defined, according to
\cite{ruderman2015}, as:
\begin{equation}
H(t)= h \, \min \Bigl[\mathrm{sign}(z + e), \max \bigl[ H(t^{-}),
\mathrm{sign} (z - e) \bigr] \Bigr], \label{eq:5}
\end{equation}
while its initial state at $t_0$ is given by
\begin{equation}
H(t_{0})= \left\{%
\begin{array}{ll}
    h \, \mathrm{sign}\bigl(z(t_{0})\bigr), & \hbox{ if } z(t_{0}) \in (-\infty, -e] \, \vee \, [e, \infty),
    \\[0.1cm]
    \left\{-h, +h\right\}, & \hbox{ otherwise.}
\end{array}%
\right. \label{eq:6}
\end{equation}
It can be seen that the hysteron has a memory of the previous
state at $t^{-}$ and keeps its value as long as $z \in (-e,e)$.

In the following, we will first assume that the relay feedback
system from Fig. \ref{fig:4} is self-sustaining, i.e., without a
disturbing factor, i.e., $y=0$. Subsequently, in sections
\ref{sec:3:sub:2} and \ref{sec:3:sub:3} we will allow for $y \neq
0$ and analyze the resulting behavior while approaching the
proposed strategy of backlash identification. Note that $y \neq 0$
corresponds to the engagement mode of the backlash, with reference
to the original system from Fig. \ref{fig:3}.

\subsection{Symmetric Unimodal Stable Limit Cycle}
\label{sec:3:sub:1}

It is well known that relay feedback systems can possess a stable
limit cycle \cite{aastrom1995}. For a single-input-single-output
(SISO) linear time invariant (LTI) system we write
\begin{eqnarray}
\label{eq:7}
  \dot{x} &=& Ax+Bu, \\
  z &=& Cx,
\label{eq:8}
\end{eqnarray}
assuming the $n \times n$ system matrix $A$ is a Hurwitz matrix,
and assuming that the input and output coupling vectors $B$ and
$C$ satisfy $CB > 0$. When a relay $u=-H[z]$ closes the feedback
loop, the state space obtains two parallel switching surfaces
which partition the state space into the corresponding cells. For
$H>0$, the system dynamics is given by $\dot{x}=Ax-Bh$, and for
$H<0$ by $\dot{x}=Ax+Bh$, while the total amplitude of the relay
switching is $2h$, see (\ref{eq:5}). If the relay in (\ref{eq:5})
is non-ideal, i.e., $e>0$, then the existence of a solution of
(\ref{eq:7}) and (\ref{eq:8}) with relay (\ref{eq:5}) in feedback
is guaranteed \cite{gonccalves2001}, and the state trajectory will
reach one of the switching surfaces for an arbitrary initial
state. The necessary condition for globally stable limit cycles of
the relay feedback system with a relay gain $h$, is given by
\begin{equation}\label{eq:9}
CA^{-1}Bh+e<0,
\end{equation}
cf. with \cite{gonccalves2001}. Otherwise, a trajectory starting
at $A^{-1}BH$ would not converge to the limit cycle. In the
following, we describe the most significant characteristics of the
stable limit cycles of the relay feedback system as shown in Fig.
\ref{fig:4}. Note that these base on the above assumptions and
conditions taken from the literature. For more details on the
existence and analysis of limit cycles in relay feedback systems
we refer to \cite{aastrom1995,johansson1999,gonccalves2001}. The
limit cycle existence condition in the control systems with
backlash and friction has also been studied in
\cite{lichtsinder2010limit}.

Considering the motor velocity dynamics (\ref{eq:1}), first
without Coulomb friction, and the relay (\ref{eq:5}) in negative
feedback, the condition for stable limit cycles (\ref{eq:9})
becomes
\begin{equation}\label{eq:10}
e < \frac{h}{d}.
\end{equation}
In fact, the velocity threshold $|e|$ should be first reachable
for a given system damping and excitation force provided during
the last switching of the relay. Therefore, (\ref{eq:10})
specifies the boundaries for parameterization of the relay, so
that to ensure the trajectory reaches one of the switching
surfaces $\pm e$ independently of the initial state. In cases
where the motor damping is not well known, its upper bound can be
assumed, thus capturing the "worst case" of an overdamped system.

Once also the Coulomb friction contributes in (\ref{eq:1}), the
condition (\ref{eq:10}) becomes a necessary but not a sufficient
one. Obviously, sufficient conditions should relate the motor
Coulomb friction coefficient $f$ to the relay, as in (\ref{eq:5}),
since both are the matched operators of one and the same argument
$\dot{x}_m$. Both are discrete switching operators, so that
\begin{equation}\label{eq:11}
h > f
\end{equation}
is required for the relay to overcome the Coulomb friction. It is
essential that the velocity sign changes while the (delayed) relay
simultaneously keeps its control value. For the boundary case $f <
h \rightarrow f$, the relay feedback system with, for instance, $H
< 0$ becomes $m \ddot{x}_m = -d \dot{x}_m + 2f$ after the velocity
sign changes from the positive to negative. Based on that, the
condition for the relay threshold results in
\begin{equation}\label{eq:12}
e < \frac{2f}{d},
\end{equation}
while following the same line of development as that for deriving
(\ref{eq:10}) from (\ref{eq:9}). Note that all three conditions
(\ref{eq:10})-(\ref{eq:12}) should hold, in order to guarantee the
existence of a stable limit cycle once the Coulomb friction is
incorporated.

\begin{figure}[!h]
\centering
\includegraphics[width=0.85\columnwidth]{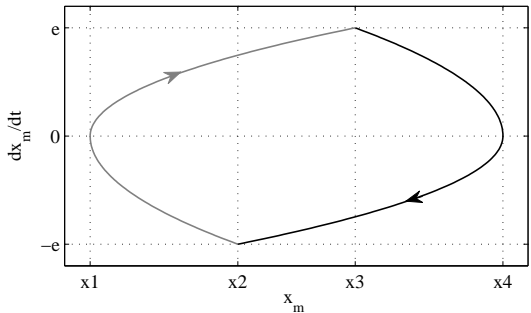}
\caption{Symmetric unimodal limit cycle.} \label{fig:5}
\end{figure}

For determining the position amplitude of the limit cycles, whose
velocity amplitude is $2e$ by definition, consider the state
trajectories of the system $m\ddot{x}_{m}+d\dot{x}_{m} + f
\mathrm{sign}(\dot{x}_{m})=-h$. The corresponding initial velocity
is $e$, immediately after the relay switches down to $-h$. It can
easily be shown that the associated phase portrait of the state
trajectories, with an initial position $x_m(0)$, can be written as
\begin{equation}\label{eq:13}
x_{m}= \frac{-0.5m\dot{x}_{m}^{2}}{d \dot{x}_{m} + h + f
\mathrm{sign}(\dot{x}_{m})} + x_m(0).
\end{equation}
Solving (\ref{eq:13}) for three characteristic velocities
$\dot{x}_{m}=\{e, 0, -e\}$, one obtains the start, maximal, and
minimal positions $x3$, $x4$, and $x2$ of the half-limit cycle, as
shown in Fig. \ref{fig:5}.
Note that in the case of $d, f = 0$, the points $x2$ and $x3$
coincide with each other. Since the symmetric relay (\ref{eq:5})
induces a symmetric unimodal limit cycle, the second half-cycle
can be obtained by mirroring the first one, derived by
(\ref{eq:13}); once across the axis connecting both switching
points at $\pm e$ and once across the orthogonal axis going
through its center (see Fig. \ref{fig:5}). The intersection of the
axes constitutes the origin of the limit cycles, which is the
manifold $\{x_{m}, | \, \dot{x}_m=0\}$. Obviously, the $x_m$
coordinate depends on the initial conditions and the system
excitation. We also recall that the limit cycle is symmetric if
$\xi(t+t^{*}) = -\xi(t)$ where $\xi(t)$ is a nontrivial periodic
solution of (\ref{eq:5})-(\ref{eq:8}) with period $2t^{\ast}$. The
limit cycle is also called unimodal when it switches only twice
per period; see \cite{gonccalves2001} for details. The symmetry
and unimodality of the limit cycle, as depicted in Fig.
\ref{fig:5}, implies that $x4-x3=x2-x1$. Therefore the total
position amplitude, denoted by $X_{\xi}$, becomes
$2(x4-x3)+(x3-x2)$. Evaluating $x2$, $x3$, $x4$ points, computed
by (\ref{eq:13}), yields
\begin{equation}\label{eq:14}
X_{\xi}=-\frac{e^{2} h \, m}{(f + h + d \,e)(f - h + d \,e)}.
\end{equation}

For determining the period of the limit cycle, the same
differential equation $m\ddot{x}_{m}+d\dot{x}_{m} + f
\mathrm{sign}(\dot{x}_{m})=-h$ is solved with respect to time,
together with the initial and final values $\dot{x}_{m}(0)=e$ and
$\dot{x}_{m}(t^{*})=-e$ respectively. This yields
\begin{equation}\label{eq:15}
   t^{*} = - \frac{m}{d} \Bigl[  \ln \Bigl(1 + \frac{ d \, e}{f-h}\Bigr)  + \ln \Bigl(\frac{ f+h}{f+h+d\,e} \Bigr) \Bigr].
\end{equation}

\subsection{Drifting Limit Cycle}
\label{sec:3:sub:2}

In the previous subsection, the conditions for a unimodal limit
cycle of the system as shown in Fig. \ref{fig:4} were derived, and
the characteristic features in terms of the displacement amplitude
and period were given by (14) and (15). We note that the limit
cycle can appear within the gap mode of backlash, (see section
II), provided $X_{\xi} < 2\beta$, so that the motor and load
remain decoupled (compare with Fig. \ref{fig:3}).

\begin{figure}[!h]
\centering
\includegraphics[width=0.7\columnwidth]{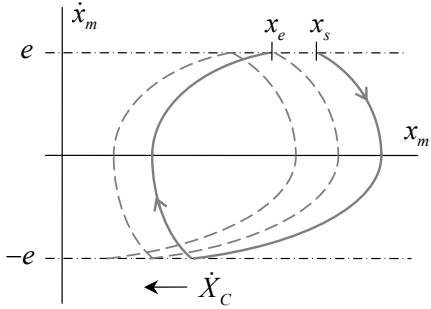}
\caption{Drifting limit cycle of amplitude-asymmetric relay
($\alpha_{-} > \alpha_{+}$)} \label{fig:6}
\end{figure}

In order to realize a drifting limit cycle, consider a modified
relay from (5) and (6) so that the amplitude becomes
\begin{equation}\label{eq:16}
h = \left\{%
\begin{array}{ll}
    \alpha_{+} h_{0},   &   \hbox{ if } \; \mathrm{sign}(H)>0, \\
    \alpha_{-} h_{0},   &   \hbox{ if } \; \mathrm{sign}(H)<0. \\
\end{array}%
\right.
\end{equation}
Here, $h_0$ is the amplitude of the underlying symmetric relay (5)
and (6), and $\alpha_{+}, \alpha_{-} \geq 1$ with $\alpha_{+} \neq
\alpha_{-}$ scales it, so as to provide an unbalanced control
effort when switching. This leads to differing acceleration and
deceleration phases of the limit cycle, so that the trajectory
does not close after one period (compare with Fig. 5), and becomes
continuously drifting as in the example shown in Fig. \ref{fig:6}.
Note that the case where $\alpha_{-} > \alpha_{+}$ is illustrated
here. As in section \ref{sec:3:sub:1}, one can solve the
trajectories between two consecutive switches separately for
$\alpha_{-} h_0$ and $\alpha_{+} h_0$, and obtain the relative
shift $X_{C} = x_e-x_s$ of the limit cycle per period as follows:
\begin{equation}\label{eq:17}
X_{C} = \frac{- e^2 h_0^2 \, m (\alpha_{-}^2 - \alpha_{+}^2) \, (f
+ d \, e) } { (- \alpha_{-}^2 h_0^2 + d^2 e^2 + 2 d e + f^2)  (-
\alpha_{+}^2 h_0^2 + d^2 e^2 + 2 d e + f^2)    } .
\end{equation}
Following the same line of developments as in section
\ref{sec:3:sub:1}, the period of the drifting limit cycle, i.e.,
the time of arrival at $x_{e}$ can be obtained as
\begin{eqnarray}
\nonumber  T_{C} &=& -\frac{m}{d} \Bigl( \ln
\frac{f+\alpha_{-}h_0}{f + d e + \alpha_{-}h_0}  + \ln \frac{f + d
e - \alpha_{-}h_0}{f -
\alpha_{-}h_0} + \\
   & + & \ln \frac{f+\alpha_{+}h_0}{f + d e +
\alpha_{+}h_0}  + \ln \frac{f + d e - \alpha_{+}h_0}{f -
\alpha_{+}h_0}   \Bigr) \label{eq:18}
\end{eqnarray}
Obviously, the average drifting velocity of the limit cycle can be
computed from (\ref{eq:17}) and (\ref{eq:18}) as $\dot{X}_{C}
\approx X_C / T_C$.
Note that (\ref{eq:17}) and (\ref{eq:18}) characterize the
drifting limit cycle which proceeds within the backlash gap
without impact with the load side. The impact behavior of drifting
limit cycles is addressed below, and that from the associated
relative load displacement point of view.

\subsection{Impact Behavior at Limit Cycles}
\label{sec:3:sub:3}

In order to analyze the impact behavior at limit cycles, i.e., the
situation where the load side is permanently shifted, as long as
it is periodically excited by the drifting limit cycle, consider
first the impact phenomenon of two colliding masses $m$ and $M$.
Note that when considering the impact and the engagement mode of
the backlash, the conditions (\ref{eq:10})-(\ref{eq:12}) are not
explicitly re-evaluated due to nontrivial (and also possibly
chaotic) solutions. However, making some smooth assumptions, we
evaluate the impact behavior at the limit cycles and show their
persistence within numerical simulations and later in experiments.

\begin{figure}[!h]
\centering
\includegraphics[width=0.85\columnwidth]{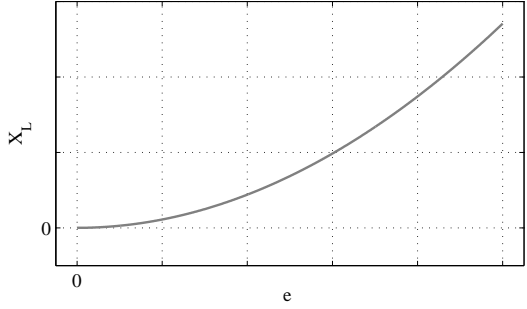}
\caption{Relative displacement of the load after the backlash
impact as a function of the relay parameter $e$ according to (22);
qualitative example.} \label{fig:7}
\end{figure}

Introducing the coefficient of restitution $0 \leq \varepsilon
\leq 1$  and denoting the velocities immediately after impact by
the superscript "$^{+}$" one can show that
\begin{eqnarray}
\label{eq:x1}
  \dot{x}_m^{+} &=& \frac{\dot{x}_L (1 - \varepsilon) M + \dot{x}_m (m- \varepsilon M)}{M+m}, \\
  \dot{x}_L^{+} &=& \frac{\dot{x}_L (M- \varepsilon m) + \dot{x}_m (1 +
  \varepsilon) m}{M+m}. \label{eq:x2}
\end{eqnarray}
The above velocity jumps directly follow from the Newton's law and
the conservation of momentum, see e.g. \cite{barbosa2002} for
details. Recall that a zero coefficient of restitution means a
fully plastic collision, while $\varepsilon = 1$ constitutes the
ideal elastic case. For the relatively low velocities at impact
(since small backlash gaps are usually assumed) and stiff
(metallic) backlash structures, we take $\varepsilon \approx 1$,
cf. with \cite{hunt1975}. Further, we assume the load velocity to
be zero and the motor velocity amplitude to be (in the worst case)
maximal, i.e., $\max |\dot{x}_m| = e$, immediately before the
impact. Due to the above assumptions, the load velocity
(\ref{eq:x2}) after impact can be determined as an upper bound:
\begin{equation}\label{eq:x3}
  \bar{\dot{x}}_L^{+} = \frac{2 \, m \, e }{M + m}.
\end{equation}
Note that (\ref{eq:x3}) constitutes an ideal case without any
structural and frictional damping at impact, whereas a real load
velocity after impact will generally be lower in amplitude than
(\ref{eq:x3}). Nevertheless, (\ref{eq:x3}) provides a reasonable
measure of the relative load motion immediately after the backlash
engages. Solving the unidirectional motion $M \ddot{x}_L + D
\dot{x}_L + F = 0$, which, for zero final velocity and the initial
velocity given by (\ref{eq:x3}), yields the relative displacement
of the load until the idle state as
\begin{equation}\label{eq:x4}
  X_L = \frac{F \, M}{D^2} \ln \Bigl( \frac{F(M+m)}{F(M+m) +2D\,e\,m} \Bigr) + \frac{2M \, e \,
  m}{D(M+m)}.
\end{equation}
Intuitively, assuming well-damped (through $F$ and $D$) load
dynamics, the load displacement $X_L$ due to a single impulse at
the impact should be relatively low. At the same time, it can be
seen that the logarithmic contribution with negative sign in the
first summand of (\ref{eq:x4}) is balanced by the linear increase
of the second summand, depending on the parameter $e$. In the sum,
the quadratic shape of $X_L$ as a function of $e$ occurs at lower
values of $e$, and it approaches a linear slope once $e$ grows
(see Fig. \ref{fig:7}). Hence, the $e$ parameter should be kept as
low-valued as possible, given the system parameters $m$, $M$, $D$
and $F$. This will ensure a displacement of the load which is as
low as possible, i.e., $X_L \rightarrow 0$, induced by the single
impulse at impact. Furthermore, low values of $e$ are required
firstly to fulfill the necessary and sufficient conditions (10)
and (12) of the limit cycle, and secondly to keep the amplitude as
low as possible and therefore within the backlash gap (see (14)).

\begin{figure}[!h]
\centering
\includegraphics[width=0.98\columnwidth]{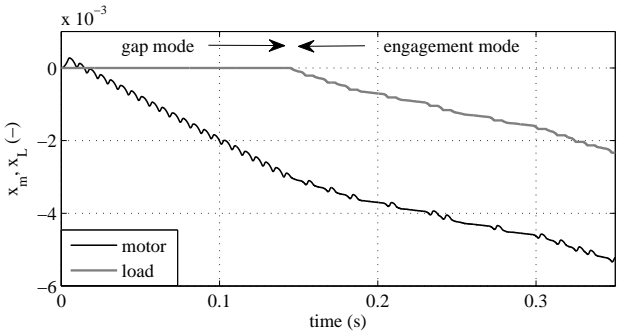}
\caption{Motor and load displacement during the gap and engagement
modes.} \label{fig:8}
\end{figure}

Obviously, the maximal load displacement (\ref{eq:x4}) will be
induced periodically by the drifting limit cycle, (see section
\ref{sec:3:sub:2}), each time the impact occurs. In fact, the
motor side, which is moving in a drifting limit cycle, produces a
sequence of impulses that continuously "push" the load side once
the instantaneous impulse can overcome the system stiction. The
resulting load motion, though periodic due to the drifting limit
cycle, can be rather chaotic, due to the amplitude and phase
shifts produced by the series of impulses. Due to the inherent
uncertainties of the Coulomb friction, viscous damping
\cite{ruderman2015b} and non-deterministic system stiction, an
exact (analytic) computation of the resulting load motion appears
to be only marginally feasible, with unsystematic errors that
reveal the predicted trajectory as less credible. At the same
time, it can be shown that for the $h$ amplitude selection to be
close to $f+F$, a continuous propulsion of the load should occur
in the direction of $\dot{X}_C$. An example of a numerical
simulation of the system (1) and (2) with backlash and an
asymmetric relay (5), (6) and (16) in feedback, is shown in Fig.
\ref{fig:8}. It can be seen that the motor displacement exhibits a
uniform drifting limit cycle during the gap mode, until it reaches
the backlash boundary and begins to interact with the load side.
It is evident that the average drifting velocity of the limit
cycle $\dot{X}_C$, corresponding to the slope in Fig. \ref{fig:8},
differs between the gap and engagement modes. This makes the
backlash boundary easily detectable based on the motor
displacement trajectory only. Note that in the numerical
simulation shown, neither damping uncertainties nor stiction of
the load is taken into account. Thus, in a real two-mass system
with backlash, a nonuniform motion during engagement mode is
expected to differ markedly from that within the backlash gap, cf.
with experiments in section \ref{sec:4}.

\section{Experimental Case Study}
\label{sec:4}

This section is devoted to an experimental case study on detecting
and identifying backlash in a two-inertia system, using only
motor-side sensing. The parameters of linear system dynamics,
i.e., inertia and damping, are first identified from the frequency
response function. The total Coulomb friction level required for
the analysis of stable limit cycles (cf. section \ref{sec:3}), is
estimated using simple unidirectional motion experiments. In
addition, the steady limit cycles previously analyzed are induced
and confirmed with experiments for unconstrained operation within
the backlash gap. The nominal backlash size was measured by means
of both motor-side and load-side encoders and used as a reference
value. The proposed backlash identification approach, as described
in section \ref{sec:3}, was followed and evaluated in the
laboratory setup. In addition, the reference method
\cite{villwock2009} was evaluated experimentally for two different
excitation conditions and compared with the proposed method.

\subsection{Laboratory Setup}
\label{sec:4:sub:1}

\begin{figure}[!h]
\centering
\includegraphics[width=0.95\columnwidth]{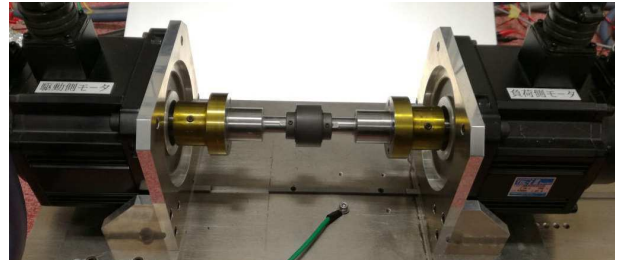}
\caption{Experimental setup with a gear coupling.} \label{fig:mb}
\end{figure}

An experimental laboratory setup (see picture in Fig.
\ref{fig:mb}), consisting of two identical motors, each with a
20-bit high-resolution encoder, was used in this study.
Note that only the motor-side encoder is utilized for the required
system identification and for evaluation of the proposed backlash
estimation method. The load-side encoder is used for reference
measurements only. In this setup, the backlash can be added or
removed by replacing the rigid coupling with the gear coupling
shown in the picture. A standard proportional-integral (PI)
current controller is implemented on-board with the motor
amplifier, with a bandwidth of 1.2 kHz. In the experiments, the
sampling frequency is set to 2.5 kHz and the control parts are
discretized using the Tustin method.

\subsection{System Identification}
\label{sec:4:sub:2}

\subsubsection{Two-Mass System Parameters}
\label{sec:4:sub:2:subsub:1}

\begin{figure}[!h]
\centering
\includegraphics[width=0.98\columnwidth]{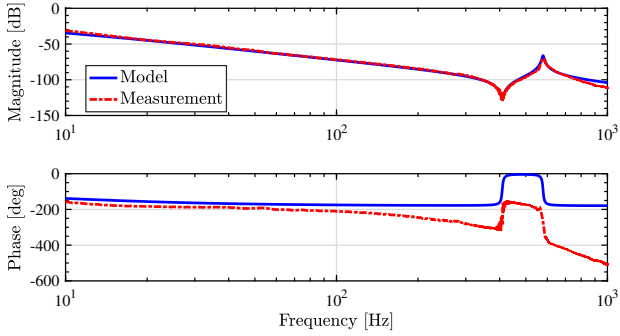}
\caption{Frequency characteristics measurements of the setup from
the motor torque to the motor-side angle versus the fitted
two-mass system model.} \label{fig:frf}
\end{figure}

The parameters of the two-mass system, i.e., $m$, $d$, $M$ and
$D$, can be identified by measuring the frequency response
function (FRF) on the motor side. It is worth emphasizing that the
identified motor inertia and damping are sufficient for analyzing
and applying the proposed backlash identification approach. Thus,
if the above parameters are available or otherwise previously
identified, the FRF-based identification discussed below will not
be required. The measured frequency characteristics, shown in Fig.
\ref{fig:frf}, disclose the antiresonance and resonance behavior.
\begin{table}[!h]
\renewcommand{\arraystretch}{1.4}
\caption{Identified parameters of two-mass system.}
\begin{center}
\begin{tabular}{|l|c|l|} \hline \hline
Motor-side inertia $m$ & 8.78e-4 &kgm$^2$ \\
\hline Motor-side damping coefficient $d$ & 6.20e-2 &Nms/rad
\\ \hline
Load-side inertia $M$ & 8.78e-4 &kgm$^2$ \\
\hline Load-side damping coefficient $D$ & 3.60e-2 &Nms/rad \\
\hline \hline
\end{tabular}
\label{tb:paras}
\end{center}
\end{table}
The parameters of the two-mass (two-inertia) system are identified
by fitting both peak regions. Here, the frequency characteristics
from the motor torque to the motor-side angle are determined for
the setup equipped with rigid coupling (without backlash). This
corresponds to the case of a nominal plant where no backlash
disturbance develops during the operation. Note that for a
sufficient system excitation, similar FRF characteristics can also
be obtained in the presence of a small backlash. Otherwise, the
basic parameters, such as lumped mass and damping of the motor and
load sides, are assumed to be available from the manufacturer's
data sheets and additional data from design correspondingly
manufacturing.

The measured FRF shows that the setup can be modeled as a two-mass
system which has an antiresonance frequency of about 409 Hz and a
resonance frequency of about 583 Hz. The fitted model response is
indicated in Fig. \ref{fig:frf} by the blue solid line, while the
measurement results are indicated by the red dashed line. A
visible discrepancy in the phase response is due to an inherent
time delay in the digital control system. However, this becomes
significant in a higher frequency range only and is therefore
neglected. The identified parameters are listed in Table
\ref{tb:paras}.

\subsubsection{Coulomb Friction}
\label{sec:4:sub:2:subsub:2}

\begin{figure}[!h]
\centering
\includegraphics[width=0.98\columnwidth]{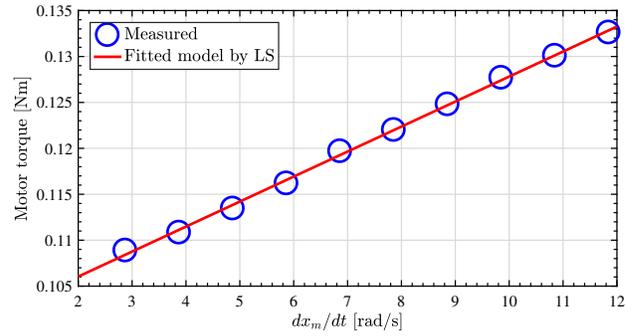}
\caption{Friction-velocity identification measurements.}
\label{fig:coulomb}
\end{figure}

The combined motor-side and load-side Coulomb friction $f+F$ is
identified by measuring the motor torque when the motor-side
velocity is controlled for constant reference values. Figure
\ref{fig:coulomb} shows the obtained torque-velocity measurements.
Blue circles indicate the measured data, captured from multiple
constant velocity drive experiments, and the red line indicates
the linear slope fitted by the least-squares method. Here, the
Coulomb friction torque of 0.1 Nm is determined from the intercept
of the fitted line with the torque axis. In the same way, the
friction-velocity curve is determined for the negative velocity
range, not shown here due to its similarity. Here, the determined
Coulomb friction value was $-0.0993$ Nm which demonstrates that
the friction behavior of the total drive is sufficiently
symmetrical around zero. An average value of $f+F = 0.0999$ Nm is
further assumed, while a very rough estimation of the motor-side
Coulomb friction is half of the total value, i.e., $f \approx
0.05$ Nm. Note that an exact knowledge of $f$ is not required to
satisfy the conditions derived in section \ref{sec:3}. Knowledge
of the total Coulomb friction coefficient $f+F$ alone is
sufficient for applying the relay feedback system.

\subsubsection{Nominal Backlash Size}
\label{sec:4:sub:2:subsub:3}

\begin{figure}[!h]
\centering
\includegraphics[width=0.98\columnwidth]{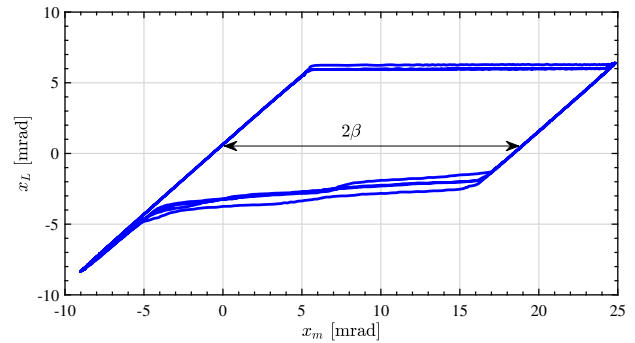}
\caption{Nominal backlash identified using both encoders.}
\label{fig:blid}
\end{figure}

For evaluation of the proposed method, the nominal backlash size
is first identified using both the motor- and load-side encoders.
Figure \ref{fig:blid} shows the measured $(x_m$,\,$x_L)$ map when
the motor-side position is open-loop controlled by a low-amplitude
and low-frequency sinusoidal wave. The detected backlash is
$2\beta = 19.05$ mrad, which can be seen as a low value, i.e.,
below one degree.

It should be emphasized that when the motor is moving forwards
after a negative direction reversal, the load side is first moving
together with the motor side, even though both are within the
backlash gap (see the lower left-hand range in Fig.
\ref{fig:blid}).
This spurious side-effect can be explained by adhesion between the
motor and load sides and the specific structure of the gear
coupling. Since the gear coupling used in the setup consists of an
internal tooth ring and two external tooth gears, it exhibits, so
to speak, a double backlash, which cannot be exactly captured by
the standard modeling assumptions previously made for a two-mass
system with backlash. It is evident that the forward transitions,
as depicted in Fig. 12 and further in Fig. 16 and Fig. 17, exhibit
some non-deterministic creep-like behavior that cannot be
attributed to backlash nonlinearity as assumed. Therefore, in
order to be able to consider the backlash effect without adhesive
disturbances, only the backward-moving phases are regarded as
suitable for evaluation.

\subsection{Steady Limit Cycle} \label{sec:4:sub:3}

\begin{figure}[!h]
\centering
\includegraphics[width=0.98\columnwidth]{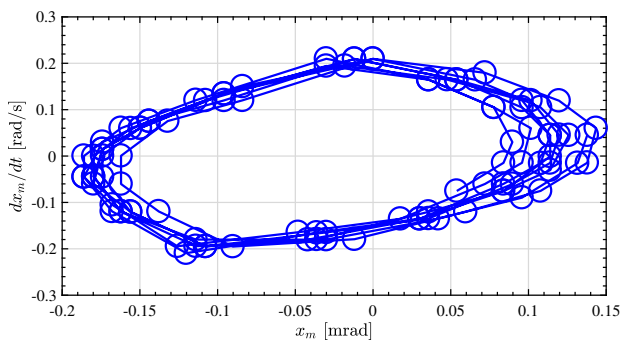}
\caption{Steady limit cycle caused by delayed relay feedback.}
\label{fig:slc}
\end{figure}

As described in section \ref{sec:3:sub:1}, the delayed relay
feedback produces steady limit cycles. The experimentally measured
limit cycle is shown in Fig. \ref{fig:slc}.
The assumed relay parameters $h=0.1$ and $e=0.1$ satisfy
(\ref{eq:10})-(\ref{eq:12}). The total amplitude $X_{\xi}$ of the
limit cycle is required to be small compared to the backlash size.
In addition, the period of the limit cycle $2 t^*$ is required to
be large enough, with respect to the system sampling rate, so as
to ensure a sufficient number of samples per period of the limit
cycle. The values computed according to (\ref{eq:14}) and
(\ref{eq:15}) are $X_{\xi} = 0.128$ mrad and $t^*=2.45$ msec
respectively. The average $X_{\xi}$ value, determined from the
example measurements shown in Fig. \ref{fig:slc}, is about 0.28
mrad. This is not surprising, since the measured steady limit
cycle is subject to the time delay of the control system so that
the actual switchings do not occur exactly at $\pm e = \pm 0.1$
rad/sec as required by the relay parameterization. Rather, they
appear at $\pm 0.2$ rad/sec, which is double the set relay
parameter value $e$. Nevertheless, the limit cycle remains stable,
almost without drifting, and displays the expected characteristic
shape, cf. with Fig. 5. Here, we note that the gear coupling with
backlash was installed, so that the steady limit cycle as shown in
Fig. \ref{fig:slc} clearly occurs within the backlash gap, without
impact with the load side. The half-period $t^*$, determined from
the measurements, corresponds to about six or seven samplings,
i.e., 2.4 to 2.8 msec, and is in accordance with that computed by
(\ref{eq:15}). Recall that knowledge of $t^*$ is significant for
deciding the sufficiency of the sampling rate with regard to the
relay control parameters.

\subsection{Motor-Side Backlash Identification}
\label{sec:4:sub:4}

The proposed backlash identification method is implemented and
evaluated experimentally. The results are evaluated for two
different sets of relay parameters. The two cases are defined as
follows. Case 1: $h_0$=0.12, $e$=0.1, $\alpha$ = 2, and Case 2:
$h_0$=0.1, $e$=0.1, $\alpha$ = 2.5. The $\alpha$ values are
selected by trial and error when operating the feedback relay
system and observing sufficient movement (drift) of the resulting
limit cycle, cf. with Fig. \ref{fig:8}. Note that a periodic
sequence for which the relay asymmetry coefficients $\alpha_{+}$
and $\alpha_{-}$ alternate, for example: $[\alpha_{+},
\alpha_{-}]$ = [2, 1] for the first five seconds followed by
$[\alpha_{+}, \alpha_{-}]$ = [1, 2] for the next five seconds, and
so on, has been applied.
\begin{figure}[!h]
\centering
\includegraphics[width=0.98\columnwidth]{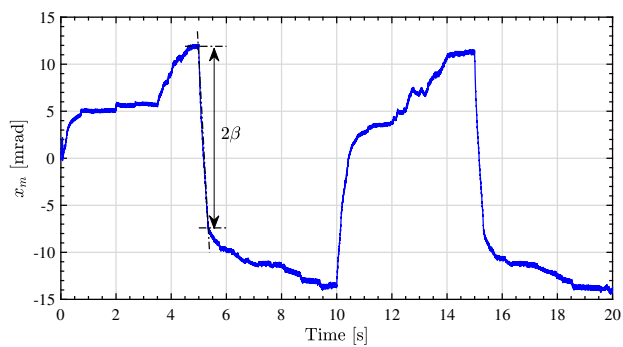}
\caption{Backlash identification using the proposed method, Case
1.} \label{fig:olbl1}
\end{figure}
This allows for changing the drift direction of the limit cycles,
and therefore exploring both the backlash gap in both directions
as well as the coupled motion beyond the gap, i.e., during the
engagement mode. Recall that only the backward-moving phases are
evaluated so as to estimate the backlash without adhesion
side-effects.

\begin{figure}[!h]
\centering
\includegraphics[width=0.98\columnwidth]{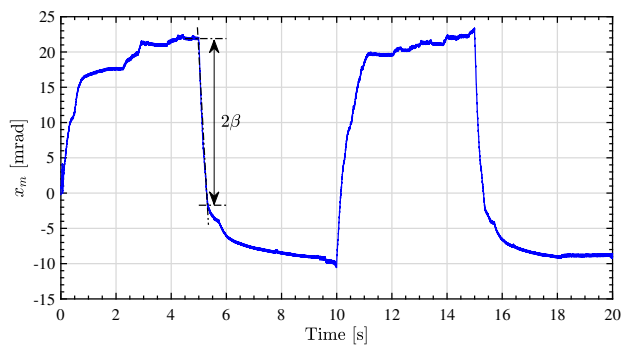}
\caption{Backlash identification using the proposed method, Case
2.} \label{fig:olbl2}
\end{figure}

The motor position responses in Case 1 and Case 2 are shown in
Fig. \ref{fig:olbl1} and Fig. \ref{fig:olbl2} respectively.
The full backlash width 2$\beta$ can be determined by the sudden
change of the motor position's slope. The slope of the motor
position is steeply and uniform within the backlash gap, since the
load is decoupled from the motor side. When the impact between
them occurs, the slope suddenly changes and becomes less steep and
also irregular. The determined 2$\beta$ values in Case 1 and Case
2 are 19.30 mrad and 23.62 mrad respectively, while the nominal
backlash width identified using the load-side encoder, as in
section \ref{sec:4:sub:2}, is 19.05 mrad. This confirms that the
proposed method can identify the backlash width using the
motor-side information only. Note that the recorded motor position
pattern is subject to various deviations between the slopes over
multiple periods which mitigates against an exact read-off of the
backlash size value (see Fig. \ref{fig:olbl1} and Fig.
\ref{fig:olbl2}). For accuracy enhancement and correspondingly
better generalization, the $2\beta$ values read-off over multiple
periods can be averaged.
\begin{figure}[!h]
\centering
\includegraphics[width=0.98\columnwidth]{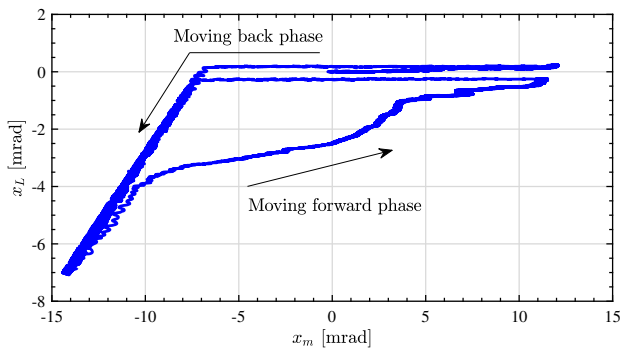}
\caption{Measured $x_L$-$x_m$ map in Case 1.} \label{fig:olblt_1}
\end{figure}
\begin{figure}[!h]
\centering
\includegraphics[width=0.98\columnwidth]{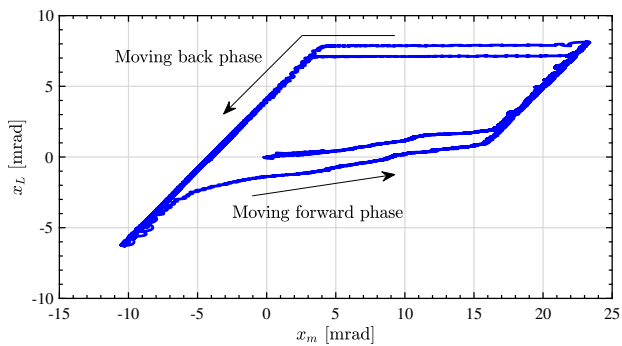}
\caption{Measured $x_L$-$x_m$ map in Case 2.} \label{fig:olblt_2}
\end{figure}
Figures \ref{fig:olblt_1} and \ref{fig:olblt_2} show the
corresponding $(x_m, x_L)$ maps, as a reference measurement, for
both experimental cases. Recall that the load-side encoder signal
is not used for backlash identification. Both figures indicate
that the limit cycles generated by the delayed relay with
asymmetric amplitudes are drifting controllably along all
transitions of the backlash. Only the forward-moving phases suffer
from adhesion side-effects, while the backwards phases coincide
exactly with the backlash shape, cf. with Fig. \ref{fig:blid}.

\subsection{Comparison with Reference Method}
\label{sec:4:sub:5}

In order to assess the performance of the proposed method, another
already established backlash identification strategy
\cite{villwock2009} has been taken as a reference method. This
method also uses the motor-side information only when identifying
the backlash in two-inertia systems. The basics of the reference
method are described below. For more details we refer to
\cite{Tustin47,villwock2009}.

The reference method applies the triangular-wave reference
velocity to the PI motor velocity controller. Considering the
controlled plant as a one-inertia system, i.e., with the lumped
parameters $m+M$ and $d+D$, the PI velocity controller has been
designed by the pole placement such that its control bandwidth is
set to 5 Hz. After the sign of the motor-side acceleration
changes, due to the triangular velocity reference, the motor side
moves back from one end to the opposite end of the backlash gap.
At the same time, the load side continues to (freely) move in the
initial direction until the backlash impact. The instant when the
motor and load sides are decoupled is defined as $t_1$ and the
instant when the motor and load sides are in contact again is
defined as $t_2$ (see Fig. 18 and Fig. 19). The reference method
identifies the backlash width by integrating the relative motor
velocity between these two instants. If the load-side viscosity is
small enough, and $t_2-t_1$ is short enough, the decrease of the
load-side velocity during $t_2-t_1$ can be neglected. Therefore,
one can assume $\dot{x}_L(t)=\hat{\dot{x}}_L=\dot{x}_m(t_1)$ for
$t_1\leq t \leq t_2$. Then, the total backlash width is estimated
as
\begin{equation}
2\beta= \left| \int \limits_{t_1}^{t_2}
\bigl(\dot{x}_L(t)-\dot{x}_m(t) \bigr)\mathrm{d}t \right| \simeq
T_s \left| \sum \limits_{k_1}^{k_2}
\bigl(\hat{\dot{x}}_L-\dot{x}_m(k)\bigr)\right|, \label{eq:conv}
\end{equation}
where $T_s$ is the sampling time and $k_1 t = t_1$ and $k_2 t =
t_2$.

\begin{figure}[!h]
\centering
\includegraphics[width=0.98\columnwidth]{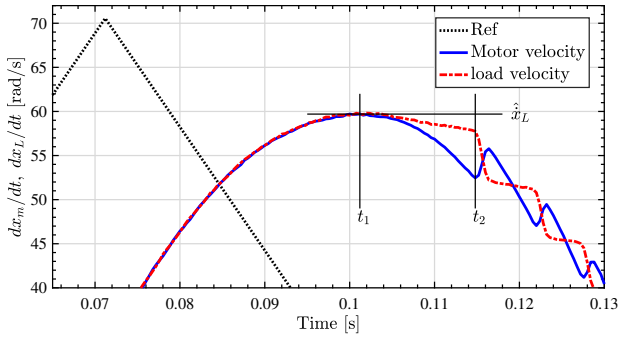}
\caption{Backlash identification using the reference method, Case
3.} \label{fig:conv_z}
\end{figure}

\begin{figure}[!h]
\centering
\includegraphics[width=0.98\columnwidth]{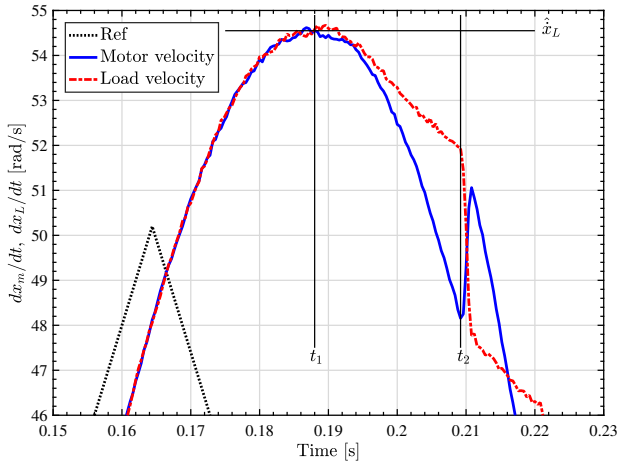}
\caption{Backlash identification using the reference method, Case
4.} \label{fig:conv_z2}
\end{figure}

The results from two different triangular waves are evaluated.
Here, two cases are defined as Case 3 -- for which the triangular
wave slope is 1400 and the period is 0.2 s, and Case 4 -- for
which the triangular wave slope is 500 and the period is 0.4 s.
Obviously, Case 3 is more "aggressive" in terms of the system
excitation, as regards the higher reference
acceleration/deceleration of the total drive. Figures
\ref{fig:conv_z} and \ref{fig:conv_z2} show the measured velocity
responses of Case 3 and Case 4 respectively. Since the
acceleration in Case 3 is larger than that in Case 4, the decrease
in $\dot{x}_L(t)$ for $t_1\leq t \leq t_2$ is smaller, which
reduces the estimation error. According to (\ref{eq:conv}), the
2$\beta$ values are calculated as 31.0 mrad and 42.6 mrad for Case
3 and Case 4 respectively.
Inherently, the reference method provides an overestimated
backlash width, as is apparent from Fig. 18 and Fig. 19. In order
to reduce the estimation error, which is mainly driven by the
assumption $\dot{x}_L(t)=\hat{\dot{x}}_L=\dot{x}_m(t_1)$, the
reference method requires larger accelerations, to make the
decoupling between the motor-side and load-side faster once the
sign of the motor acceleration changes. Therefore, the reference
method inherently requires more "aggressive" system excitations
and generally larger load inertias to allow for a free,
correspondingly decoupled, load motion phase. At the same time,
higher and uncertain damping and Coulomb friction values will lead
to higher errors when estimating $2\beta$ according to
(\ref{eq:conv}).

The backlash identification results for all four above cases, i.e.
for the proposed method and reference method, are summarized in
Table \ref{tab:summery}, while the nominal backlash size was
measured as 19.05 mrad.
\begin{table}[!h]
  \renewcommand{\arraystretch}{1.5}
  \caption{Backlash identification results summary}
  \label{tab:summery}
  \begin{center}
  \begin{tabular} {||p{1.5cm}|p{1.5cm}||p{1.5cm}|p{1.5cm}||}
  \hline
  Case 1     &   Case 2   &  Case 3    &   Case 4   \\
  \hline
  19.3 mrad  &   23.6 mrad   &  31.0 mrad      &   42.6 mrad       \\
  \hline
  \end{tabular}
  \end{center}
  \normalsize
\end{table}

\section{Conclusions}
\label{sec:5}

A new approach for identifying backlash in two-mass systems is
proposed, with the principal advantage of using the motor-side
sensing only, which is a common practice for various machines and
mechanisms. The method is based on the delayed relay in feedback
of the motor velocity, which allows for inducing the stable limit
cycles of amplitudes significantly lower than the backlash gap.
The limit cycles can then be operated as drifting while being
controlled by an asymmetric relay amplitude. The analyzed period
of limit cycles, cf. (\ref{eq:15}), imposes certain boundary on
applicability of the method and that in relation to the sampling
time and possible time delays of sensing and actuating in the
relay feedback loop. Another factor inherently limiting the
proposed method is the sensing resolution of the motor-side,
correspondingly the accuracy with which the relative velocity used
for relay feedback can be measured.

We provide a detailed analysis of steady and drifting limit cycles
for two-mass systems with backlash, and derive the conditions for
the minimal set of system parameters which can be easily
identified. The experimental evaluation of the method is performed
on a test bench with two coupled motors, where the rigid coupling
is replaceable by one with a small backlash gap of about one
degree. The known identification method reported in
\cite{villwock2009} is taken as a reference for comparison.
Several advantages of the proposed method, in terms of less
aggressive system excitations and no need for high-speed motions
of the overall two-mass system, are shown and discussed along with
the experimental results.

\begin{acknowledgment}
This work has received funding from the European Union's Horizon
2020 research and innovation programme (H2020-MSCA-RISE-2016)
under the Marie Sklodowska-Curie grant agreement No 734832.
\end{acknowledgment}

%
% ...

\bibliographystyle{asmems4}

\bibliography{references}

\end{document}